\documentclass[english, 12pt, a4paper]{article}
\usepackage[utf8]{inputenc}
\usepackage{authblk}

\usepackage[numbers,compress]{natbib}
\usepackage{graphicx}
\usepackage{amsmath}
\usepackage{amssymb}
\usepackage[ruled,vlined]{algorithm2e}

\begin{document}
\title{Ptychographic phase-retrieval by proximal algorithms}
\author{Hanfei Yan \\hyan@bnl.gov}
\affil{National Synchrotron Light Source II, Brookhaven National Laboratory, Upton, NY 11973, USA}
\date{\today}

\maketitle

\begin{abstract}
\normalsize
We derive a set of ptychography phase-retrieval iterative engines based on proximal algorithms originally developed in convex optimization theory, and discuss their connections with existing ones. The use of proximal operator creates a simple frame work that allows us to incorporate the effect of noise from a maximum-likelihood principle. We focus on three particular algorithms, namely proximal minimization, alternating direction method of multiplier and accelerated proximal gradient, and benckmark their performance with numerical simulations and experimental x-ray data. Among them, accelerated proximal gradient shows superior performance in terms of both accuracy and convergence rate for a noisy dataset.           
\end{abstract}

\section{Introduction}
Ptychography is a powerful scanning imaging technique that utilizes advanced mathematical tools to retrieve the missing phase information of the wave-field from a sequence of intensity measurements \citep{RN1313,RN1103,RN347}. The attraction of this technique comes from its capability of recovering both the complex-valued probe and object functions, a blind deconvolution, as well as its ability of breaking the resolution barrier set by the focusing optics. It gains increasing popularity in recent years for its robustness in practice and was used successfully for many imaging applications in different fields \citep{RN1104,RN1285,RN1283,RN933,RN1314,RN1315}. The major challenge of this technique resides in the fact that the mathematical problem is non-convex and ill posed. Particularly for real-world problems, the experimental data always contains noise and other types of errors, therefore finding a solution optimized globally is extremely difficult, if not completely impossible. A great deal of efforts have been devoted to developing robust ptychographic iterative engines, either based on alternating-projection or gradient-descent methods \citep{RN1088,RN347,RN1101,RN1302,RN1295,RN1298,RN1296,RN1299,RN1303,RN1297}. More complex algorithms that can handle mixed states \citep{RN1307}, positioning errors \citep{RN1308,RN1309}, diffraction effect \citep{RN1316,RN1318} and instability of the probe \citep{RN1312} were developed as well.  

For convex optimization problems, a class of algorithms called proximal algorithms have been studied extensively \citep{RN1288}. They turn out to be well-suited for constrained, large-scale and distributed optimization problems, and ptychography falls into this category. In these techniques, the solving process is divided into sub-problems involving the evaluation of the proximal operator, which usually has a closed-form solution. Inspired by these developments, here we propose to combine proximal algorithms and Wirtinger derivatives to create a simple frame work for solving ptychography problems, where either alternating-projection or gradient-descent algorithms can be derived straightforwardly. 

We want to emphasize that proximal algorithms are originally developed for convex problems dealing with real-valued numbers, and ptychography problems are non-convex and involves complex-valued numbers. The Wirtinger derivative allows us to use the common rules for differentiation known from real-valued analysis \citep{RN1291}, so the developed solving techniques in proximal algorithms can be readily applied. We show some previously reported ptychography algorithms can be derived in this frame work. Here we do not attempt to provide a rigorous theoretical proof of the overall convergence, but rather offer a heuristic for future work by demonstrating the effectiveness of these algorithms with numerical simulation and real-world applications. As have been shown that many phase-retrieval algorithms can find their counterparts in convex optimization theory \citep{RN1319}, so can ptychography. In this paper, we focus on ptychographic reconstruction with noise and round-off errors, a common problem encountered in all measurements. Three maximum-likelihood (ML) based algorithms, namely proximal minimization (PM), alternating direction method of multiplier (ADMM) and accelerated proximal gradient (APG) are derived and benchmarked with both simulation and experiment data. Among them, APG, which has not been reported before, exhibits a superior performance. Although current model only considers noise and round-off errors, it is not difficult to extend it to a more complex case.      

This paper is organized as follows. In the model section, we first build the mathematical model for the optimization problem in ptychography, and then discuss how different solving algorithms can be derived from proximal operators. Connections with existing techniques are discussed. Two statistical models for the noise, intensity Poission and amplitude Gaussian, are considered throughout the paper. In the numerical simulation section, we benchmark performance of the derived algorithms at different signal-to-noise ratios, and discuss the optimal conditions for convergence and accuracy in respective cases. In the experimental data section, PM, ADMM and APG are tested with an experimental dataset taken with x-rays. The result confirms that APG outperforms the other two and achieves the state-of-the-art performance.              

\section{Model}
 In a ptychographic scan, a probe, $p$, impinges on different parts of an object, $o$. The transmitted wave is assumed to be simply a product of the probe and object function, and propagates to a farfield detector. The wavefield at the object plane is linked to the wavefield at detector plane by Fourier transform. Because a physical detector only measures intensity (or amplitude in other words), the phase of the wavefield is lost in an experiment. A phase-retrieval algorithm intends to recover this set of missing information based on the measured amplitude under certain conditions. Assuming that we collect farfield diffraction patterns at $K$ different positions, We now have two constraints for the measured data to satisfy. One is in real space (sample plane). The wavefield has to be written as a product of a probe and an object function with known translations. The other is in reciprocal space (detector plane). The modulus of its Fourier transform has to agree with the measurement. 
\begin{equation}
     y_i = |F[p(\mathbf{r})o(\mathbf{r}-\mathbf{r}_i)]|,\quad 
     i = 1,2,...K,
\end{equation}
where $y_i$ is the measured amplitude of the ith image and $F$ is the two-dimensional Fourier transform operator. In Eq. (1), another set of known information is the probed position, $\mathbf{r}_i$. The goal here is to find complex-valued functions $p$ and $o$ that satisfy both the amplitude and translation constraints. We discretize the problem and rewrite Eq.(1) into a vector form by concatenating an image along its column directions,
\begin{gather}
    \mathbf{y}_i=|\mathbf{x}_i|,\quad
    \mathbf{x}_i = \mathbf{F} \mathbf{P} \mathbf{S}_i \mathbf{o}.
\end{gather}
Here the lower-case bold letter represents a column vector and a capital bold one corresponds to a matrix. The absolute operator is element-wise. In the above equation, $\mathbf{y}_i \in \mathbb{R}^{N \times 1}$ is the measured amplitude, $\mathbf{F} \in \mathbb{C}^{N \times N}$ is the Fourier transform matrix, $\mathbf{p} \in \mathbb{C}^{N\times 1}$ is the probe vector, $\mathbf{P} = \text{diag}(\mathbf{p}$) is a diagonal matrix, $\mathbf{S}_i \in \mathbb{R}^{N \times M}$ is a sparse matrix containing only zeros and ones that selects the illuminated elements of the object vector, $\mathbf{o} \in \mathbb{C}^{M \times 1}$, at the ith probed position. The resulting vector is $\mathbf{o}_i \in \mathbb{C}^{N \times 1}$. Problem described by Eq. (2) can be turned into an unconstrained optimization problem,
\begin{equation}
    \underset{\mathbf{x}_{i}}{\text{min}} \quad f(\mathbf{x}_i)+g(\mathbf{x}_i),
\end{equation}
where $f$ and $g$ are indicator functions corresponding to the two constraints respectively,
\begin{gather}
    f(\mathbf{x}_i) = \left\{ \begin{array}{cc}
  0 &\mbox{ if $\mathbf{x}_i \in \mathbf{dom}f $} \\
  \infty &\mbox{otherwise}
       \end{array}, \quad \mathbf{dom}f = \{\mathbf{x}_i \in \mathbb{C}^{N \times 1} \mid |\mathbf{x}_i|= \mathbf{y}_i \},\right.\nonumber\\
  g(\mathbf{x}_i) = \left\{ \begin{array}{cc}
  0 &\mbox{ if $\mathbf{x}_i \in \mathbf{dom}g $} \\
  \infty &\mbox{otherwise}
       \end{array}, \quad \mathbf{dom}g = \{\mathbf{x}_i \in \mathbb{C}^{N \times 1} \mid \mathbf{x}_i=\mathbf{F}\mathbf{P}\mathbf{S}_i\mathbf{o}\}.\right.
\end{gather}

\subsection{Alternating Projection}
The classical method of solving the above problem is alternating projection algorithm (AP) that projects the solution into two domains,
\begin{gather}
    \mathbf{z}_i^{t+1}=\Pi_f (\mathbf{x}_i^t), \nonumber\\
    \mathbf{x}_i^{t+1}= \Pi_g (\mathbf{z}_i^{t+1}).
\end{gather}
Here $\Pi_f$ and $\Pi_g$ are Euclidean projection operators. By iteratively projecting into these two domains, we hope that an initial guess can converge to a solution fulfilling both constraints. 

A proximal operator of a function $f$ is defined  by,
\begin{equation}
    \mathbf{prox}_{\lambda f}(\mathbf{v})=\underset{\mathbf{x}}{\text{argmin}}(f(\mathbf{x})+\frac{1}{\lambda}\|\mathbf{x}-\mathbf{v}\|^2_2).
\end{equation}
Here we use $1/\lambda$ instead of $1/2\lambda$ for convenience because we need to deal with complex-valued variables. For the two indicator functions in Eq. (3) their corresponding proximal operators are simplified to,
\begin{eqnarray}
    \mathbf{prox}_f (\mathbf{v}_i)&=& \underset{\mathbf{x}_{i}\in \mathbf{dom}f}{\text{argmin}} \quad \sum_{i=1}^{K} \|\mathbf{x}_i-\mathbf{v}_i\|^2_2 \nonumber \\
    &=&\Pi_f(\mathbf{v}_i),\nonumber \\
     \mathbf{prox}_g (\mathbf{v}_i)&=& \underset{\mathbf{x}_i\in \mathbf{dom}g}{\text{argmin}} \quad \sum_{i=1}^{K} \|\mathbf{x}_i-\mathbf{v}_i\|^2_2 \nonumber \\
     &=&\Pi_g(\mathbf{v}_i).
\end{eqnarray}
 Based on the definition, these proximal operators are just Euclidean projections. Therefore, the classical AP method can also be written as,
 \begin{gather}
     \mathbf{z}_i^{t+1}=\mathbf{prox}_f(\mathbf{x}_i^{t}),\nonumber\\
     \mathbf{x}_i^{t+1}= \mathbf{prox}_g(\mathbf{z}_i^{t+1}).
 \end{gather}
 
 The solution to the first projection is simply to replace the amplitude of $\mathbf{v}_i$ with $\mathbf{y}_i$ while retain its phase,
 \begin{eqnarray}
     \Pi_f(\mathbf{v}_i) = \text{diag}(\mathbf{y}_i)\mathbf{\vartheta}(\mathbf{v}_i),\quad
     \mathbf{\vartheta}(\mathbf{v}_i) = \left\{\begin{array}{cc}
          \mathbf{v}_i/|\mathbf{v}_i| &\mbox{$\mathbf{v}_i \neq 0$}  \\
          0&\mbox{otherwise} 
     \end{array}\right.
 \end{eqnarray}
 The divide is an element-wise operation. The Solution to the second projection can be obtained by many different ways. The probe and object functions can be updated sequentially \citep{RN1299}, collectively \citep{RN347} or jointly \citep{RN1298}. Here we choose the usual collective update,
 \begin{gather}
     \mathbf{o}=(\sum_{i=1}^{K}\mathbf{S}_i^H\mathbf{P}^H\mathbf{P}\mathbf{S}_i)^{-1} (\sum_{i=1}^{K}\mathbf{S}_i^H\mathbf{P}^H\mathbf{F}^H\mathbf{v}_i),\nonumber\\
     \mathbf{p}=(\sum_{i=1}^{K}\mathbf{O}_i^H\mathbf{O}_i)^{-1} (\sum_{i=1}^{K}\mathbf{O}_i^H\mathbf{F}^H\mathbf{v}_i),\nonumber\\
     \mathbf{O}_i = \text{diag}(\mathbf{S}_i\mathbf{o}),\nonumber\\
     \Pi_g(\mathbf{v}_i) = \mathbf{F}\mathbf{P}\mathbf{S}_i\mathbf{o},
 \end{gather}
 where superscript $H$ denotes conjugate and transpose operation. $\Pi_g$ defines another projection satisfying the second constraint. Specifically, we back-propagate the wavefield to the sample plane using inverse Fourier transform. Then we update probe and object functions collectively based on the probed positions. One can run the iteration once or multiple times for high accuracy. Lastly, we replace the wavefield at the sample plane by the product of the probe and object functions and propagate it to the detector plane. This method is often referred as error reduction (ER) algorithm.
 
 From a statistical point of view, if we assume a Gaussian likelihood function of the amplitude and use its negative log as our cost function, we arrive at,
\begin{equation}
    \mathcal{L} = \sum_{i=1}^{K} {\| \mathbf{y}_i-|\mathbf{F}\mathbf{P}\mathbf{S}_i\mathbf{o}| \|_2^2}.
\end{equation}
Their Wirtinger derivatives with respect to the probe and object are,
\begin{gather}
    \frac{\partial \mathcal{L}}{\partial \mathbf{o^*}} = \sum_{i=1}^{K} {(\mathbf{F} \mathbf{P}\mathbf{S}_i)}^H [ \mathbf{x}_i-\text{diag}(\mathbf{y}_i)\vartheta(\mathbf{x}_i)],\nonumber\\
    \frac{\partial \mathcal{L}}{\partial \mathbf{p^*}} = \sum_{i=1}^{K} (\mathbf{F} \mathbf{O}_i)^H [\mathbf{x}_i-\text{diag}(\mathbf{y}_i)\vartheta(\mathbf{x}_i)].
\end{gather}
At a stationary point the derivatives have to be zero,
\begin{gather}
    \sum_{i=1}^{K}(\mathbf{S}_i^H \mathbf{P}^H \mathbf{P} \mathbf{S}_i) \mathbf{o} - \sum_{i=1}^{K}(\mathbf{S}_i^H \mathbf{P}^H \mathbf{F}^H) \text{diag}(\mathbf{y}_i)\vartheta(\mathbf{x}_i) = 0,\nonumber\\
    \sum_{i=1}^{K} (\mathbf{O}_i^H \mathbf{O}_i) \mathbf{p} - \sum_{i=1}^{K} \mathbf{O}_i^H \mathbf{F}^H \text{diag}(\mathbf{y}_i)\vartheta(\mathbf{x}_i) = 0.
\end{gather}
They can be solved iteratively by a fixed-point algorithm that seeks a fix point of the equation, $\mathbf{z}=q(\mathbf{z})$. In this case, the unknown variables are $\mathbf{o}$ and $\mathbf{p}$. If they are solved in sequence we arrive at,
\begin{eqnarray} 
    \mathbf{o}^{t+1} &=& [\sum_{i=1}^{K} (\mathbf{P}^{t}\mathbf{S}_i)^H \mathbf{P}^{t} \mathbf{S}_i]^{-1} \nonumber \\
    &&[\sum_{i=1}^{K} (\mathbf{P}^{t}\mathbf{S}_i)^H \mathbf{F}^H \text{diag}(\mathbf{y}_i)\vartheta(\mathbf{x}_i^t)],\nonumber\\
    \mathbf{p}^{t+1} &=& (\sum_{i=1}^{K} {\mathbf{O}_i^{t+1}}^H \mathbf{O}_i^{t+1})^{-1} \nonumber \\
    &&[\sum_{i=1}^{K} {\mathbf{O}_i^{t+1}}^H \mathbf{F}^H \text{diag}(\mathbf{y}_i)\vartheta(\mathbf{x}_i^t)],\nonumber\\
    \mathbf{x}_i^{t+1} &=& \mathbf{F} \mathbf{P}^{t+1} \mathbf{S}_i \mathbf{o}^{t+1}.
\end{eqnarray}
This is no different from the classical ER algorithm shown in Eq. (10). Therefore, we can also interpret ER algorithm for ptychography as some sort of fix-point algorithm that seeks the stationary point of the amplitude Gaussian likelihood function with respect to $\mathbf{p}$ and $\mathbf{o}$.     

\begin{algorithm}[H]
\SetAlgoLined
\KwData{$\mathbf{y}_i \in \mathbb{R}^{N \times 1}$, $\mathbf{S}_i \in \mathbb{R}^{N \times M}$, $i = 1,2,...K$}
\KwResult{probe function $\mathbf{p}$ and object function $\mathbf{o}$}
 initialization: $\mathbf{p}^0$, $\mathbf{o}^0$, $\mathbf{x}_i^0$, $t_{max}$\;
 \Repeat{$t>t_\text{max}$}{
 $\mathbf{z}_i^{t+1}=\text{diag}(\mathbf{y}_i)\mathbf{\vartheta}(\mathbf{x}_i)$\;
 $\mathbf{o}^{t+1}=\text{update\_o}(\mathbf{p}^t,\mathbf{z}_i^{t+1})$\;
 $\mathbf{p}^{t+1}=\text{update\_p}(\mathbf{o}^{t+1},\mathbf{z}_i^{t+1})$\;
 $\mathbf{x}_i^{t+1}=\mathbf{F}\text{diag}(\mathbf{p}^{t+1})\mathbf{S}_i\mathbf{o}^{t+1}$\;
 }
 \caption{ER}
\end{algorithm}

\subsection{Alternating Direction Method of Multiplier}
For real-world phase-retrieval problems, ER is known to suffer from slow convergence and stagnation issues. A far more robust and popular algorithm is ADMM. One special variant of its form is Douglas-Rachford splitting method, also known as difference map (DM), which is widely used for ptychography reconstructions. ADMM is usually derived from argumented Lagrangian method. In the framework of proximal algorithms, it can be written in a very concise form. An in-depth discussion of ADMM can be found in the monography by Boyd et al. \citep{RN1289}, and its application for ptychography were reported in previous publications \citep{RN1302,RN1297}. Recently it was applied for joint ptycho-tomography reconstruction \citep{RN1305}. Thus, here we skip the derivation process. We change our optimization problem [Eq. (3)] slightly,
\begin{eqnarray}
    &\underset{\mathbf{x}_{i},\mathbf{z}_i}{\text{min}} \quad f(\mathbf{x}_i)+g(\mathbf{z}_i),&\nonumber \\
    &\text{subject to} \quad \mathbf{x}_i=\mathbf{z}_i.&
\end{eqnarray}
For two indicator functions defined in Eq.(4), the ADMM algorithm is,
\begin{eqnarray}
    &\mathbf{x}_i^{t+1} = \mathbf{prox}_f(\mathbf{z}_i^{t}-\mathbf{u}_i^{t}),&\nonumber\\
    &\mathbf{z}_i^{t+1} = \mathbf{prox}_g(\mathbf{x}_i^{t+1}+\mathbf{u}_i^{t}),&\nonumber\\
    &\mathbf{u}_i^{t+1}=\mathbf{u}_i^{t}+\mathbf{x}_i^{t+1}-\mathbf{z}_i^{t+1}.&
\end{eqnarray}
If we define a new variable $\mathbf{w}_i^t = \mathbf{z}_i^t+\mathbf{u}_i^t$ and substitute it into Eq. (16), we arrive at,
\begin{eqnarray}
    &\mathbf{x}_i^{t+1} = \mathbf{prox}_f(2\mathbf{z}_i^{t}-\mathbf{w}_i^{t}),&\nonumber\\
    &\mathbf{z}_i^{t+1} = \mathbf{prox}_g(\mathbf{x}_i^{t+1}+\mathbf{w}_i^{t}-\mathbf{z}_i^t),&\nonumber\\
    &\mathbf{w}_i^{t+1}=\mathbf{w}_i^{t}+\mathbf{x}_i^{t+1}-\mathbf{z}_i^{t}.&
\end{eqnarray}
Variable $\mathbf{x}_i^{t+1}$ is not independent and can be replaced. Rearrange terms and use Euclidean projection operators derived in Eqs. (9) and (10), we have, 
\begin{eqnarray}
    &\mathbf{z}_i^{t}=\Pi_g(\mathbf{w}_i^{t}),&\nonumber\\
    &\mathbf{w}_i^{t+1}=\mathbf{w}_i^{t}+\Pi_f(2\mathbf{z}_i^{t}-\mathbf{w}_i^t)-\mathbf{z}_i^t.&
\end{eqnarray}
This is the well-known DM algorithm \citep{RN347}. 

For a noisy dataset, DM is known to have stability problem because it attempts to find a solution with its amplitude exactly equal to the measured value. A simple remedy is to replace the indicator function, $f$, with a negative log likelihood function, $\mathcal{L}$. In such a case, $\mathbf{x}_i$ update in Eq. (16) is modified to,
\begin{eqnarray}
    \mathbf{x}_i^{t+1} &=& \mathbf{prox}_{\lambda \mathcal{L}}(\mathbf{z}_i^{t+1}-\mathbf{u}_i^t), \nonumber\\
    \mathbf{prox}_{\lambda \mathcal{L}}(\mathbf{v}_i) &=&  \frac{\text{diag}(\vartheta{(\mathbf{v}_i)})}{2(1+\lambda)} \nonumber\left\{ \begin{array}{cc}
  2(\lambda\mathbf{y}_i+|\mathbf{v}_i|), &\mbox{\text{G}} \\
  |\mathbf{v}_i|+(\sqrt{|\mathbf{v}_i|^2+4\lambda(1+\lambda)\mathbf{y}_i^2}), &\mbox{P}
       \end{array} \right. \\
    &=&\text{E}_\text{MAP}(\lambda,\mathbf{y}_i,\mathbf{v}_i)
    \end{eqnarray}  
Here `G' and `P' refers to amplitude Gaussian and intensity Poisson, respectively. One may notice that the update scheme for $\mathbf{x}_i$ is now parameter-dependent. From Bayes' theorem, the proximal operator can be interpreted as maximum-a-posterior (MAP) probability estimate, where the prior probability follows a normal distribution. The parameter, $\lambda$, controls how close the new update should be to its prior value, and plays an important role in determining the performance of the algorithm. We will have a more detailed dissuasion in the following section.    

\begin{algorithm}[H]
\SetAlgoLined
\KwData{$\mathbf{y}_i \in \mathbb{R}^{N \times 1}$, $\mathbf{S}_i \in \mathbb{R}^{N \times M}$, $i = 1,2,...K$}
\KwResult{probe function $\mathbf{p}$ and object function $\mathbf{o}$}
 initialization: $\mathbf{p}^0$, $\mathbf{o}^0$, $\mathbf{z}_i^0$, $\mathbf{u}_i^0$, $\lambda$, $\beta$, $\delta$, $t_{max}$\;
 \Repeat{$t>t_{max}$}{
 $\mathbf{x}_i^{t+1}=\text{E}_\text{MAP}(\lambda, \mathbf{y}_i,\mathbf{z}_i^t-\mathbf{u}_i^t)$\;
 $\mathbf{o}^{t+1}=\text{update\_o}(\mathbf{p}^t,\mathbf{x}_i^{t+1}+\mathbf{u}_i^t)$\;
 $\mathbf{p}^{t+1}=\text{update\_p}(\mathbf{o}^{t+1},\mathbf{x}_i^{t+1}+\mathbf{u}_i^t)$\;
 $\mathbf{z}_i^{t+1}=\mathbf{F}\text{diag}(\mathbf{p}^{t+1})\mathbf{S}_i\mathbf{o}^{t+1}$\;
 $\mathbf{u}_i^{t+1}=\mathbf{u}_i^t+\mathbf{x}_i^{t+1}-\mathbf{z}_i^{t+1}$\;
 \If{$\|\mathbf{u}^{t+1}-\mathbf{u}^{t}\|_2/\|\mathbf{u}^{t+1}\|_2<\delta$}{
 $\lambda = \beta\lambda$\;
 }
 }  
 \caption{mADMM}
\end{algorithm}

\subsection{Proximal Minimization}
The fixed point of a proximal operator is also the minimizer of the original function. This leads to a simple proximal iterative algorithm,
\begin{equation}
    \mathbf{x}_i^{t+1}=\mathbf{prox}_{\lambda\mathcal{L}+g}(\mathbf{x}_i^t).
\end{equation}
For the ptychography problem considered in this paper, we define,
\begin{eqnarray}
    \mathbf{prox}_{\lambda\mathcal{L}+g} (\mathbf{v}_i) &=& \underset{\mathbf{x}_i\in \mathbf{dom}g}{\text{argmin}} \quad \left. \mathcal{L}(\mathbf{x}_i)+g(\mathbf{x}_i)+\sum_{i=1}^{K}\frac{1}{\lambda}{\|\mathbf{x}_i-\mathbf{v}_i\|^2_2}\right.\nonumber\\
    &=& \underset{\mathbf{p},\mathbf{o}}{\text{argmin}} \quad\left. \mathcal{L}(\mathbf{p},\mathbf{o})+\sum_{i=1}^{K}\frac{1}{\lambda}{\|\mathbf{F}\mathbf{P}\mathbf{S}_i\mathbf{o}-\mathbf{v}_i\|^2_2}\right.
\end{eqnarray}
Again, we can interpret the update as a MAP estimate. The difference from ADMM discussed above is that $\mathbf{x}_i$ and $\mathbf{z}_i$ are forced to be equal here. In ADMM, the splitted variables belong to their individual domains and are not necessary the same. Similar to the derivation of Eq. (14), we make two-step update,
\begin{eqnarray}
    &\mathbf{z}_i^{t+1} =  \text{E}_\text{MAP}(\lambda,\mathbf{y}_i,\mathbf{x}_i^t),& \nonumber\\
    &\mathbf{x}_i^{t+1} = \Pi_g(\mathbf{z}_i^{t+1}).&
\end{eqnarray}
Compared to ER, the only difference is that the measured amplitude is replaced with a MAP-estimated value. This method is first proposed by Katkovnik et al. \citep{RN1295}. Here we show it is equivalent to PM algorithm and provide an alternative perspective.  

\begin{algorithm}[H]
\SetAlgoLined
\KwData{$\mathbf{y}_i \in \mathbb{R}^{N \times 1}$, $\mathbf{S}_i \in \mathbb{R}^{N \times M}$, $i = 1,2,...K$}
\KwResult{probe function $\mathbf{p}$ and object function $\mathbf{o}$}
 initialization: $\mathbf{p}^0$, $\mathbf{o}^0$, $\mathbf{x}_i^0$, $\lambda$, $t_{max}$\;
 \Repeat{$t>t_\text{max}$}{
 $\mathbf{z}_i^{t+1}=\text{E}_\text{MAP}(\lambda, \mathbf{y}_i,\mathbf{x}_i^t)$\;
 $\mathbf{o}^{t+1}=\text{update\_o}(\mathbf{p}^t,\mathbf{z}_i^{t+1})$\;
 $\mathbf{p}^{t+1}=\text{update\_p}(\mathbf{o}^{t+1},\mathbf{z}_i^{t+1})$\;
 $\mathbf{x}_i^{t+1}=\mathbf{F}\text{diag}(\mathbf{p}^{t+1})\mathbf{S}_i\mathbf{o}^{t+1}$\;
 }
 \caption{PM}
\end{algorithm}

\subsection{Accelerated Proximal Gradient}
Let's consider the optimization problem,
\begin{eqnarray}
    \underset{\mathbf{x}_{i}}{\text{min}} \quad \mathcal{L}(\mathbf{x}_i)+g(\mathbf{x}_i).
\end{eqnarray}
Wirtinger derivative allows us to derive the gradient of the real-valued likelihood function with respect to the complex-valued variable $\mathbf{x}_i$,
\begin{gather}
  \nabla_{\mathbf{x}_i^*}\mathcal{L}=\left\{\begin{array}{cc}
  \mathbf{x}_i-\text{diag}(\mathbf{y}_i)\vartheta(\mathbf{x}_i), &\quad\mbox{\text{G}} \\
  \mathbf{x}_i-\text{diag}(\frac{\mathbf{x}_i}{|\mathbf{x}_i|^2+\varepsilon})(\mathbf{y}_i^2+\varepsilon),&\quad\mbox{P}
       \end{array} \right.
\end{gather}
Here $\varepsilon$ is a small real-valued constant introduced to avoid the discontinuity at zero, as suggested in Ref \citep{RN1297}. The negative of the gradient is also the steepest descent direction of the function. The proximal gradient algorithm is,
\begin{eqnarray}
    \mathbf{x}_i^{t+1}=\mathbf{prox}_g(\mathbf{x}_i^t-{\lambda^t}\nabla_{\mathbf{x}_i^*}\mathcal{L}(\mathbf{x}_i^t))
\end{eqnarray}
Here $\lambda^t$ is a positive step size that can vary at each iteration. We use a simple method to determine its value \citep{RN1300}. The step size remains the same unless the following condition is violated
\begin{gather}
    \mathbf{z}_i=\mathbf{prox}_g(\mathbf{x}_i^t-\lambda^t\nabla_{\mathbf{x}_i^*}\mathcal{L}(\mathbf{x}_i^t)),\nonumber\\
    \mathcal{L}(\mathbf{z}_i)\leq\mathcal{Q}_{\lambda^t}(\mathbf{z}_i,\mathbf{x}_i^t),\nonumber\\
    \mathcal{Q}_{\lambda^t}(\mathbf{z}_i,\mathbf{x}_i^t)=\mathcal{L}(\mathbf{x}_i^t)+\sum_{i=1}^{K}2\text{Re}(\nabla_{\mathbf{x}_i^*}\mathcal{L}(\mathbf{x}_i^t)^H(\mathbf{z}_i-\mathbf{x}_i^t))+\frac{1}{\lambda^t}\|\mathbf{z}_i-\mathbf{x}_i^t\|_2^2,
\end{gather}
In such a case, we reject the update and multiply the step size by a factor $\beta\in (0,1)$. The process is repeated until the above condition is satisfied and then the iteration is completed, $\mathbf{x}_i^{t+1}=\mathbf{z}_i$. The proximal gradient algorithm can be understood from a point of view of localized optimization. For completeness, we give an explanation due to Beck and Teboulle \citep{RN1300}. The function, $\mathcal{Q}_{\lambda^t}(\mathbf{z}_i,\mathbf{x}_i^t)$, is an upper bound to $\mathcal{L}(\mathbf{x}_i^t)$ that is tight at $\mathbf{x}_i^t$, i.e. $\mathcal{Q}_{\lambda^t}(\mathbf{x}_i^t,\mathbf{x}_i^t)=\mathcal{L}(\mathbf{x}_i^t)$ and $\mathcal{Q}_{\lambda^t}(\mathbf{z}_i,\mathbf{x}_i^t)\geq\mathcal{L}(\mathbf{z}_i)$, provided that $\lambda^t\in(0,L]$, where $L$ is a Lipschitz constant of $\nabla_{\mathbf{x}_i^*}\mathcal{L}$. This function can be considered as an first-order approximation to $\mathcal{L}$ with a regularization term. We may rewrite it as,
\begin{equation}
    \mathcal{Q}_{\lambda^t}(\mathbf{z}_i,\mathbf{x}_i^t)=\mathcal{L}(\mathbf{x}_i^t)+\sum_{i=1}^{K}\frac{1}{\lambda^t}\|\mathbf{z}_i-(\mathbf{x}_i^t-\lambda^t\nabla_{\mathbf{x}_i^*}\mathcal{L})\|_2^2-\lambda^t\|\nabla_{\mathbf{x}_i^*}\mathcal{L}\|_2^2
\end{equation}
In the vicinity of $\mathbf{x}_i^t$, we replace the original optimization problem with an approximate one,
\begin{equation}
    \underset{\mathbf{z}_i}{\text{min}}\quad\mathcal{Q}_{\lambda^t}(\mathbf{z}_i,\mathbf{x}_i^t)+g(\mathbf{z}_i)
\end{equation}
Dropping constant terms in Eq. (27) does not affect the solution to Eq. (28). As a result we arrive at,
\begin{eqnarray}
    &&\underset{\mathbf{z}_i}{\text{argmin}}\quad \sum_{i=1}^{K}\frac{1}{\lambda^t}\|\mathbf{z}_i-(\mathbf{x}_i^t-\lambda^t\nabla_{\mathbf{x}_i^*}\mathcal{L})\|_2^2+g(\mathbf{z}_i)\nonumber\\
    &&=\mathbf{prox}_g(\mathbf{x}_i^t-\lambda^t\nabla_{\mathbf{x}_i^*}\mathcal{L})
\end{eqnarray}
Consequently, we can interpret each iteration as a proximal operator of $g$ along the steepest decent direction of $\mathcal{L}$, as the name proximal gradient suggests. By definition,
\begin{equation}
    \mathcal{L}(\mathbf{x}_i^{t+1})+g(\mathbf{x}_i^{t+1})\leq\mathcal{Q}_{\lambda^t}(\mathbf{x}_i^{t+1},\mathbf{x}_i^t)+g(\mathbf{x}_i^{t+1})\leq\mathcal{Q}_{\lambda^t}(\mathbf{x}_i^t,\mathbf{x}_i^t)+g(\mathbf{x}_i^{t}).
\end{equation}
In this particular case $g$ is an indicator function, thus, 
\begin{equation}
    g(\mathbf{x}_i^{t})=g(\mathbf{x}_i^{t+1})=0
\end{equation}
We can simplify the inequality as,
\begin{equation}
    \mathcal{L}(\mathbf{x}_i^{t+1})\leq\mathcal{Q}_{\lambda^t}(\mathbf{x}_i^{t+1},\mathbf{x}_i^t)\leq\mathcal{Q}_{\lambda^t}(\mathbf{x}_i^t,\mathbf{x}_i^t)=\mathcal{L}(\mathbf{x}_i^{t}).
\end{equation}
Therefore, each iteration decent the negative log likelihood function meanwhile satisfying the constraint $g$. For a faster convergence, the accelerated version of the proximal gradient method which includes an additional extrapolation step can be used,
\begin{gather}
    \mathbf{w}_i^{t} = \mathbf{x}_i^t+\omega^{t}(\mathbf{x}_i^{t}-\mathbf{x}_i^{t-1}),\nonumber\\
    \mathbf{x}_i^{t+1}=\mathbf{prox}_g(\mathbf{w}_i^{t}-\lambda^t\nabla_{\mathbf{w}_i^*}\mathcal{L}(\mathbf{w}_i^{t})),\nonumber\\
    \omega^t = \frac{t}{t+3}.
\end{gather}

We note that Xu et al. \citep{RN1296} recently proposed accelerated Wirtinger flow (AWF) method for ptychography, which stems from the popular Wirtinger flow (WF) algorithm for phase-retrieval problems \citep{RN1293}. It shares some similarity with APG. However, they differ fundamentally in many aspects. AWF can be considered as a steepest decent method with a constant step size, while APG here is a projected gradient method with a varying step size. We may consider APG as a hybid algorithm combining gradient descent and projection methods. We first decent the cost function in reciprocal space with a small step, and then project it to the domain in real space which satisfies the translation constraint. The update is completed only when such a move would make the cost function smaller. For a Gaussian amplitude likelihood function, if $\lambda^t$ is equal to one, as can be seen the updating scheme is no different from ER algorithm. Therefore, we can choose one as the initial value of the step size.       

\begin{algorithm}[H]
\SetAlgoLined
\KwData{$\mathbf{y}_i \in \mathbb{R}^{N \times 1}$, $\mathbf{S}_i \in \mathbb{R}^{N \times M}$, $i = 1,2,...K$}
\KwResult{probe function $\mathbf{p}$ and object function $\mathbf{o}$}
 initialization: $\mathbf{p}^0$, $\mathbf{o}^0$, $\mathbf{x}_i^0$, $\lambda^0=1$, $\beta$, $t_{max}$\;
 \Repeat{$t>t_\text{max}$}{
 $\omega^t=\frac{t}{t+3}$\;
 $\mathbf{w}_i^t=\mathbf{x}_i^t+\omega^t(\mathbf{x}_i^t-\mathbf{x}_i^{t-1})$\;
 \Repeat{$\lambda^t$ is sufficiently small}{
 $\mathbf{z}_i=\mathbf{w}_i^t-{\lambda^t}\nabla_{\mathbf{x}_i^*}\mathcal{L}(\mathbf{w}_i^t)$\;
 $\mathbf{o}^{t+1}=\text{update\_o}(\mathbf{p}^t,\mathbf{z}_i)$\;
 $\mathbf{p}^{t+1}=\text{update\_p}(\mathbf{o}^{t+1},\mathbf{z}_i)$\;
 $\mathbf{z}_i=\mathbf{F}\text{diag}(\mathbf{p}^{t+1})\mathbf{S}_i\mathbf{o}^{t+1}$\;
 \eIf{$\mathcal{L}(\mathbf{z}_i)\leq\mathcal{Q}_{\lambda^t}(\mathbf{z}_i,\mathbf{w}_i^t)$}{
 \Return $\mathbf{x}_i^{t+1}=\mathbf{z}_i$\;
 }
 {
 $\lambda^t=\lambda^t\beta$\;
 }
 }
 $\lambda^{t+1}=\lambda^{t}$\;
 \If{$\lambda^{t+1}<\delta$}{
 $\lambda^{t+1}=\lambda^0$\;
 }
 }
 \caption{APG}
\end{algorithm}

\section{Numerical simulation}
In this section we will compare the performance of different methods using simulation data. The test object function is shown in Fig. 1. The image  `Cameraman' is used as its amplitude and `Barbara' as its phase. The pixel size of the image is assumed to be 5 nm. A probe of size 37 nm is produced by a Fresnel Zone plate and a special fermat scan that follows the equation (in polar coordinates) \citep{RN1022}, 
\begin{equation}
    r_i = c\sqrt{i},\quad\theta_i =2.4i,\quad i=1,2...K.
\end{equation}
is performed to illuminate the different parts of the sample. $c$ is chosen to be 20 nm (4 pixels) and a total of 1261 far-field diffraction patterns are collected. The maximum detector intensity is scaled to the range of $10^2-10^6$ and a Poisson noise is added to each pixel accordingly. The "measured" intensity contains two different types of errors. One is the round-off error since the measured intensity are integers. This approximation reduces the dynamical range of the signal. Particularly in the detector region where the intensity drops below 0.5 count, they are all set to zero. Two is the Poisson noise, which adds background fluctuation to the signal. To be more quantitative, we calculate the signal-to-noise ratio (SNR) of the intensity as,
\begin{equation}
    \text{SNR}=10\text{log}_{10}(\frac{\sum_{i=1}^{K}\|\mathbf{\hat{y}}_i^2\|_2^2}{\sum_{i=1}^{K}\|\mathbf{y}_i^2-\mathbf{\hat{y}}_i^2\|_2^2}),
\end{equation}
 where $\mathbf{\hat{y}}_{i}$ are the ground-truth values. 
 
 We first study the case with SNR = 32.25 dB (max. detector pixel intensity = $10^4$ counts). The reconstruction results obtained from different algorithms discussed in the preceding section are shown in Fig. 1. All of ML-based methods yield a high-quality reconstruction with nearly indistinguishable difference. If we pay close attention on the reconstructed amplitude, a very faint cross-talk from the phase image can be seen in the background. In contrast, the phase reconstruction doesn't show any visible artifacts. As a comparison, we also plot results from the non-statistical algorithm, DM. Strong cross-talk in the reconstructed amplitude image can be observed. In addition, the reconstructed phase image is less sharp than the others and contains some visible artifacts. This suggests that even at this level of signals, one may still need to use ML-based algorithm to achieve the best result.    
 
\begin{figure}[!ht]
\centering
\includegraphics[scale=0.5]{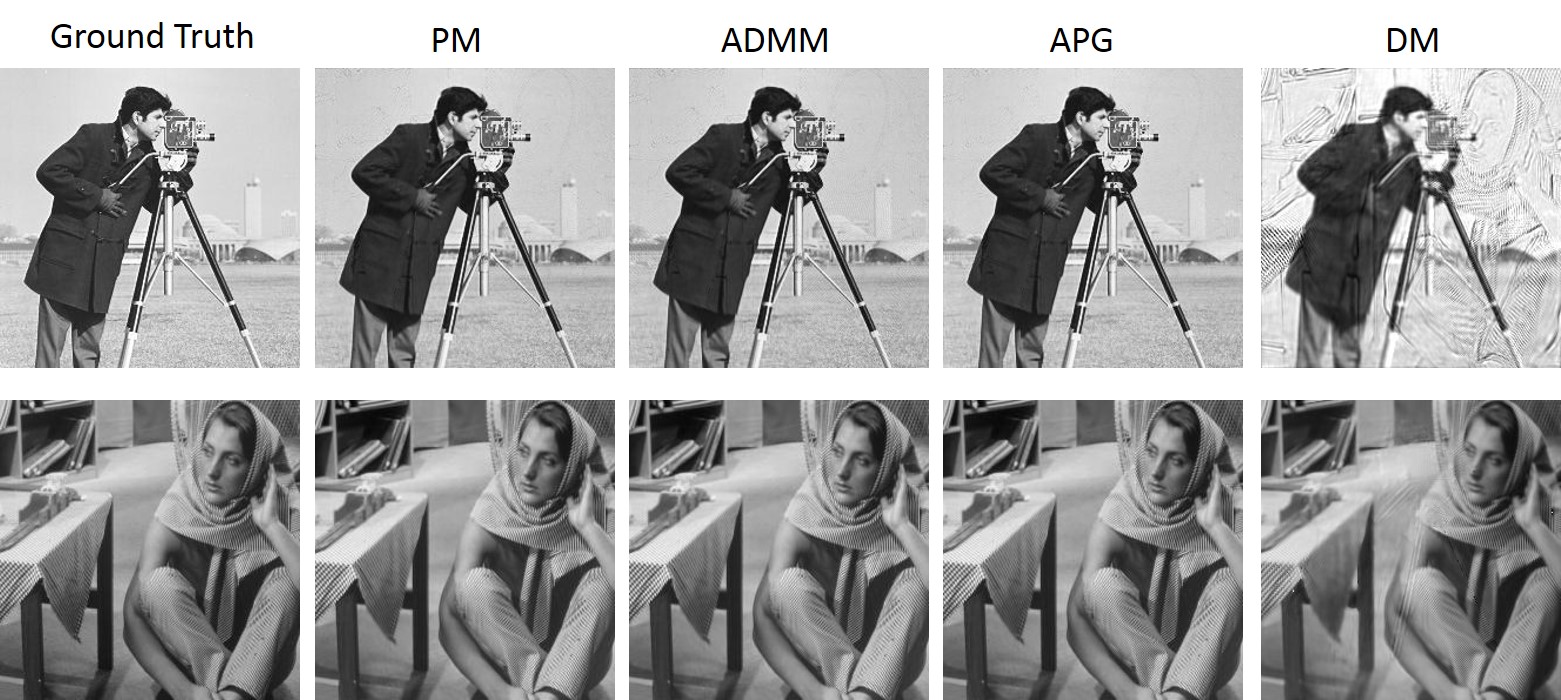}
\caption{Comparison of reconstruction results obtained from different algorithms with a noisy dataset. The maximum number of photons received at one pixel of the detector is scaled to $10^4$ counts, and a Poisson noise is added accordingly. This corresponds to SNR = 32.25 dB for the collected diffraction patterns. Top panel: amplitude image. Bottom panel: phase image.}
\label{fig:figure1}
\end{figure}

 To quantify the error for a systematic study, we use a root mean square error (RMSE) defined as, \begin{equation}
     \text{RMSE}=\frac{\|\mathbf{\hat{o}}-a\mathbf{o}\|_2}{\|\mathbf{\hat{o}}\|_2},
 \end{equation}
 where $\mathbf{\hat{o}}$ is the ground-truth and $a$ is a complex-valued constant to account for the ambiguity in ptychography reconstruction. The assessment under different conditions is presented in Fig. 2. For PM (Fig. 2a), its convergence rate is $\lambda$-dependent, while the resultant RMSE is not; they all converge to the same value. The Poisson model leads to a smaller RMSE, which is not a surprise since a Poisson noise is added. When the amplitude Gaussian is used, PM is no better than ER. As we discussed earlier, ER can also be considered as a ML-based method.   
\begin{figure}[!ht]
\centering
\includegraphics[scale=0.8]{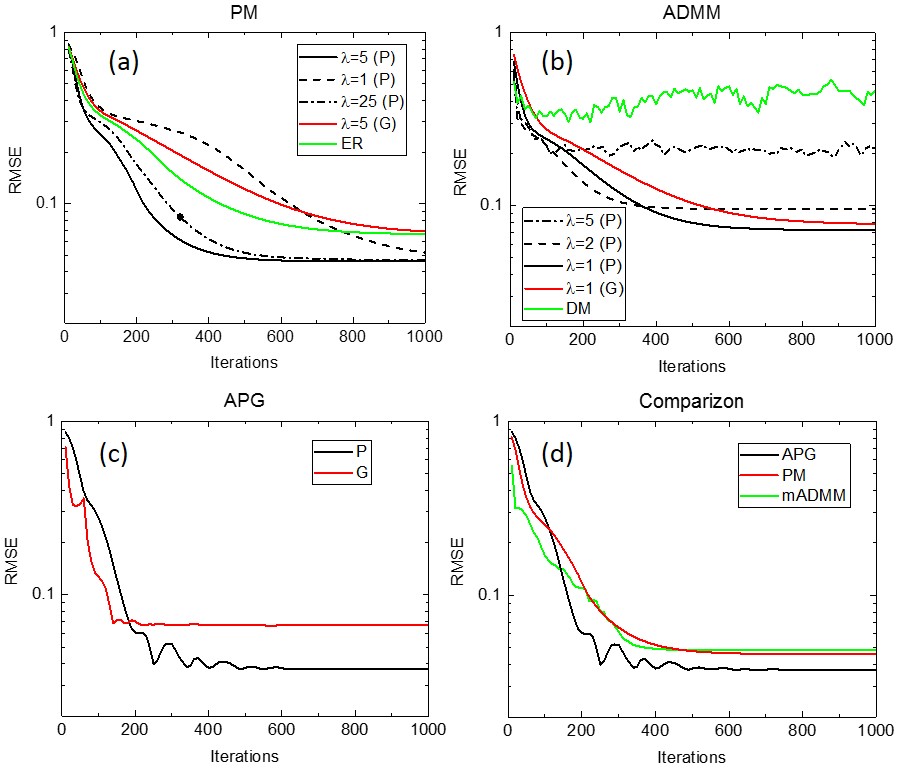}
\caption{RMSE variations as a function of iteration numbers with (a) PM, (b) ADMM, (c) APG algorithms under different conditions. A comparisons across algorithms is shown in (d), where a multi-stage strategy is employed for ADMM (named mADMM) to achieve the best result, $\delta = 10^{-5}, \beta = 0.7$. The simulated dataset used for reconstruction is the same with that in Fig. 1.}
\label{fig:figure2}
\end{figure}

 For ADMM (Fig. 2b), we observe that a larger value of $\lambda$ leads to a faster convergence. In the limit of infinity, ML-based ADMM becomes DM, which shows the fastest convergence rate at the beginning. However, a large value of $\lambda$ can cause a stability issue, which can be seen in the plot. In this case when $\lambda>5$, they do not tend to converge to a stable solution after initial fast convergence, but rather fluctuate as iteration goes. Also, a large $\lambda$ results in a higher RMSE. Therefore the solution is less accurate. This suggests a multi-stage strategy for ADMM to optimize both the convergence rate and accuracy. We can start with a large $\lambda$ for fast convergence. When a stable solution is reached ($\|\mathbf{u}^{t}-\mathbf{u}^{t-1}\|/\|\mathbf{u}^{t}\|<\text{threshold}$), we reduce the value of $\lambda$ by multiplying it with a constant $\beta\in(0,1)$, use the solution obtained from the last stage as the initial guess and continue the iteration. To distinguish it from a regular ADMM algorithm, we call it mADMM thereafter, where `m' refer to multi-stage.  
 
 For APG (Fig. 2c), we always choose the initial value of $\lambda^0$ as one and it is adjusted automatically in the update. Similar to the cases in PM and ADMM, intensity Poisson model outperforms amplitude Gaussian. It is worth noting that when the initial guess of the probe is bad, $\lambda^t$ can quickly becomes very small, particularly for intensity Poisson model. This will cause a stagnation problem. A remedy is to reset $\lambda^t$ to its initial value when it becomes too small, but a price to pay is more computation time. 
 
 To assess performance across algorithms, in Fig. 2d we plot their RMSE variations as a function of iteration number. For a fair comparison, all start with a disk-like probe and a square object as the initial guess. Poisson model is used in all algorithms. Among them, APG has the best performance. Not only its overall convergences rate is higher, but also the resultant RMSE is smaller.
 
\begin{figure}[!ht]
\centering
\includegraphics[scale=0.6]{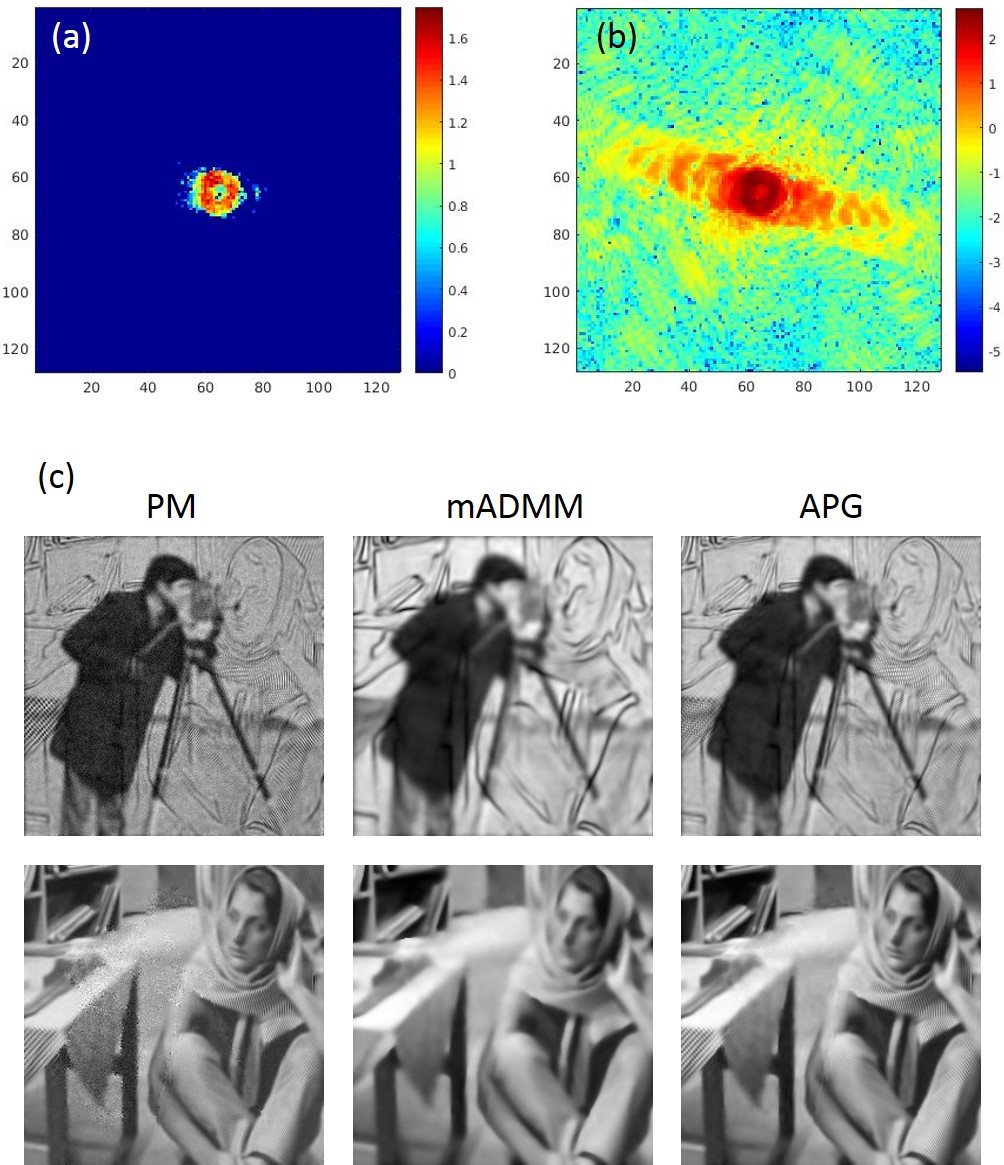}
\caption{A typical diffraction pattern of the noisy dataset (SNR=12.77) (a) and the corresponding error-free one (b). Intensity plot in logarithmic scale. (c) reconstruction amplitude and phase images using PM ($\lambda = 1$), mADMM ($\lambda=1,\delta = 10^{-5},\beta=0.7$) and APG ($\delta = 0.1,\beta=0.5$) algorithms with the noisy dataset seen in (a).}
\label{fig:figure3}
\end{figure}

 With current SNR, though quantitative analysis shows the difference in the reconstruction, visually the results look almost the same (Fig. 1). In a more extreme case, we consider a maximum detector intensity of 100 counts (SNR = 12.77). Compared to the previous case, the diffraction intensity is reduced by 100-fold. Fig. 3a shows a typical diffraction pattern with limited counts in logarithmic scale, and a comparison with the ground-truth (Fig. 3b). The error-free data has a dynamical range over seven orders of magnitude, while that for the limited counts is reduced to less than two due to the round-off to integers. The added noise makes the situation even worse. Such a noisy dataset poses a significant challenge to ptychographic reconstruction. We present in Fig. 3c the reconstructed results obtained from PM, APG and mADMM algorithms. Because DM does not converge at all in this case, its result is not shown here. Again a disk-like probe and a square object as the initial guess are used. All three algorithms lead to a converged solution. As is clear, their results however differ considerably from algorithm to algorithm. PM recovers high-frequency features well, but the reconstructed images look more grainy. There are also strong cross-contamination between the amplitude and the phase images. In contrast, mADMM yields smooth images, but the high-frequency details are lost. APG balances the two well and produces the best overall results. The quality of the obtained phase image is still very acceptable, without having visible artifacts and losing too much details.    

\begin{figure}[!ht]
\centering
\includegraphics[scale=1]{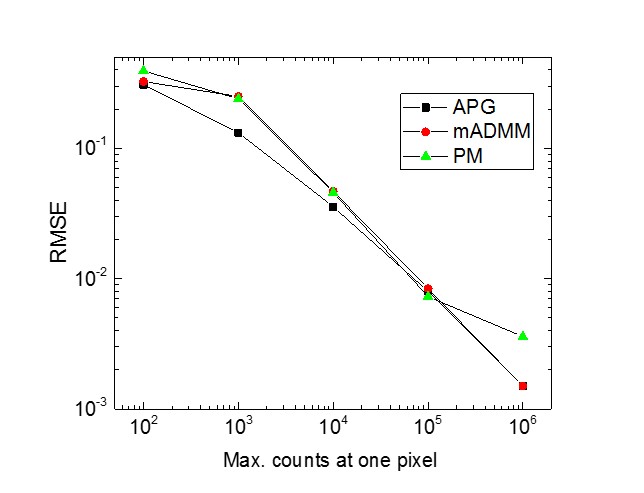}
\caption{RMSE variations at different levels of signal strength using PM, mADMM and APG algorithms.}
\label{fig:figure4}
\end{figure}
 
 Fig. 4 depicts the achieved RMSE of the reconstructed complex image at various levels of signal. In general, the log-log plot shows a close-to-linear relationship, suggesting that they all follow a power law approximately. In most circumstances, APG outperforms the other two. Unlike PM and mADMM algorithms for which the value of $\lambda$ has to be chosen accordingly with the noise level to achieve the best performance, there is no need to tune any parameter for APG when SNR changes.     
 
 \section{Experimental Data}   
 The simulation data only take into account round-off errors and Poisson noise. The real-world problem can be much more complicated. In order to assess the robustness of these algorithms, we perform reconstruction on an experimental dataset that was taken at the hard x-ray nanoprobe beamline of National Synchrotron Light Source II, Brookhaven National Laboratory. The nanobeam with a size of $\sim$13 $\text{nm}^2$ at 12 keV was produced by two crossed multilayer Laue lenses. Details about the experimental setup can be found elsewhere \citep{RN1276}. The sample consists of cubic Au nanoparticles with a size of 50 nm deposited on a Si substrate. They form an ordered array with gaps between them as small as 10 nm. The sample was placed at a downstream position with a distance of 25 um to the focal plane. A fermat scan with $c=20$ nm [Eq. (34)] was performed to avoid periodic aliasing effect, and a total of 792 frames were collected. The diffraction pattern collected on a far-field pixel-array detector (Merlin, Quantum Detectors) has over $4\times 10^4$ maximum detector counts at one pixel. 
 
 In Fig. 5 reconstructed complex-valued images with different algorithms are presented. For a fair comparison, they all start with the same initial guess of the probe function, which is obtained by inverse Fourier transforming the measured far-field amplitude and then propagating 25 um to the sample plane. In other words, the initial guess assumes a lens with no phase aberration.  Because DM does not converge to a stable solution, we have to choose an intermediate reconstruction result that looks the best. Nevertheless, the phase image is a bit noisy, and the amplitude part is barely recognizable. PM yields a smoother result, but both the phase and amplitude exhibit some ghost image around the boundary of the array. mADMM produces a further improved result, but the ghost image can still be seen, particularly in the amplitude image. Not surprisingly, the best result is achieved with APG. It has no apparent artifacts seen in the reconstruction. We can clearly resolve the shape of individual 50-nm nanoparticles and their sharp edges in the phase image, even though the amount of the phase variation of one layer is in the order of $\sim$0.05 radian. Because the sample has a very low absorption contrast ($\sim$1.7\%), the amplitude image usually is too fuzzy to be useful. However in the one obtained from APG we can still recognize the particle array. For this dataset, we conclude that APG leads to a reconstruction result with overall image quality noticeably better than that of others. 
 
\begin{figure}[!ht]
\centering
\includegraphics[scale=0.62]{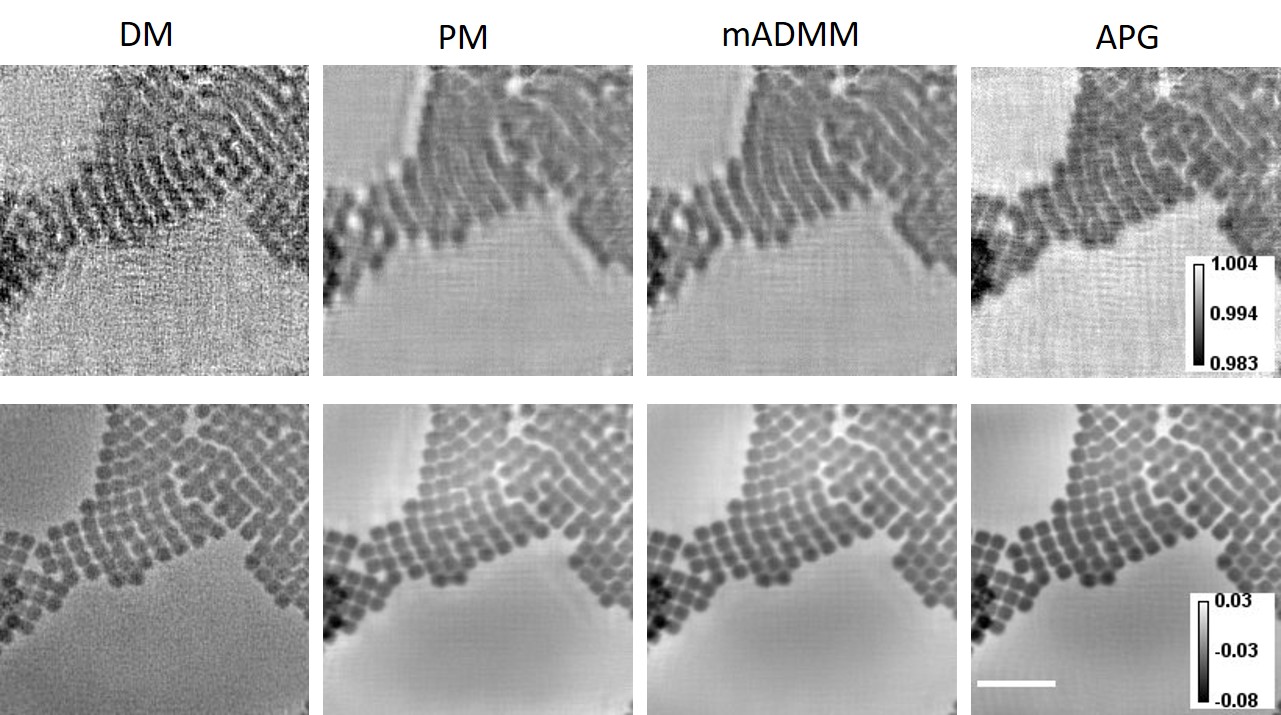}
\caption{Ptychography reconstruction of a Au nanoparticle array using different algorithms. Scale bar is 250 nm. Top panel: amplitude. Bottom panel: phase.}
\label{fig:figure5}
\end{figure}
 
 There are a few remarks we'd like to point out. Among all the algorithms tested in this paper, DM takes the least number of iterations to arrive at a plausible solution, particularly when the initial guess of the probe is far from the ground truth. However, with the presence of noise DM can become unstable and diverge. On the contrary, ML-based algorithms converge slower, but are stable. The reason is that their iteration processes usually involves a MAP-based sub-optimization step which requires the updated solution to be close to its prior value. As a result, a good guess of the initial probe is more important for these ML-based algorithms. In addition, for the experimental data tested in this case we do not see visible difference between Poisson and Gaussian models. This suggests that the difference in reconstruction between two statistical models diminishes as intensity increases. 
 
 \section{Conclusion} 
 In summary, we presented several solving techniques for ptychographic imaging derived in the framework of proximal algorithms. The separable nature of the proximal operator makes it well-suited for dealing with large-scale ptychography reconstruction problems where its evaluation can be parallelized. The optimization problem is usually divided into sub-optimization steps involving proximal operators for which often a closed-form solution can be found. Therefore, the problem becomes more tractable. We derived ER, PM, ADMM, DM and APG algorithms and benchmarked their performance with noisy datasets. Among them, APG depicted the best reconstruction result not only in numerical simulation but also in experiment.   
 
 In the current work, we only consider a noise model and round-off errors. In the same frame work, it is not difficult to enable modes to deal with partial coherence \citep{RN1307} and bluring effect in fly-scan \citep{RN1152,RN1153}. In such cases, contribution from different modes will add up incoherently and the likelihood function has to be modified accordingly. We can also consider to incorporate more constraints on probe or object function. For example, adding a regularization term with denoiser has shown suppressed noise in the reconstructed object function \citep{RN1295,RN1311}. With an additional constraint on the probe, it was demonstrated that the periodic aliasing effect seen in grid scans could be mitigated \citep{RN1297}. There is still a lot of room for improvement, and they will be the future work. 
 
\section{Acknowledgements}
The author thanks X. Huang for a fruit discussion on ptychography algorithms and F. Lu for providing Au nanoparticles. This research used beamline 3ID of the National Synchrotron Light Source II, a U.S. Department of Energy (DOE) Office of Science User Facility operated for the DOE Office of Science by Brookhaven National Laboratory under Contract No. DE-SC0012704. 

\bibliographystyle{unsrt}
\bibliography{references}
\end{document}